\documentclass[12pt]{article}
\newlength{\dinwidth}
\newlength{\dinmargin}
\usepackage{graphicx}          % PostScript graphics, extended syntax

\setkeys{Gin}{keepaspectratio} % We will specify a width AND a
\begin{document}

\def\ap#1#2#3   {{\em Ann. Phys. (NY)} {\bf#1} (#2) #3.}   
\def\apj#1#2#3  {{\em Astrophys. J.} {\bf#1} (#2) #3.} 
\def\apjl#1#2#3 {{\em Astrophys. J. Lett.} {\bf#1} (#2) #3.}
\def\app#1#2#3  {{\em Acta. Phys. Pol.} {\bf#1} (#2) #3.}
\def\ar#1#2#3   {{\em Ann. Rev. Nucl. Part. Sci.} {\bf#1} (#2) #3.}
\def\cpc#1#2#3  {{\em Computer Phys. Comm.} {\bf#1} (#2) #3.}
\def\err#1#2#3  {{\it Erratum} {\bf#1} (#2) #3.}
\def\ib#1#2#3   {{\it ibid.} {\bf#1} (#2) #3.}
\def\jmp#1#2#3  {{\em J. Math. Phys.} {\bf#1} (#2) #3.}
\def\ijmp#1#2#3 {{\em Int. J. Mod. Phys.} {\bf#1} (#2) #3}
\def\jetp#1#2#3 {{\em JETP Lett.} {\bf#1} (#2) #3.}
\def\jpg#1#2#3  {{\em J. Phys. G.} {\bf#1} (#2) #3.}
\def\mpl#1#2#3  {{\em Mod. Phys. Lett.} {\bf#1} (#2) #3.}
\def\nat#1#2#3  {{\em Nature (London)} {\bf#1} (#2) #3.}
\def\nc#1#2#3   {{\em Nuovo Cim.} {\bf#1} (#2) #3.}
\def\nim#1#2#3  {{\em Nucl. Instr. Meth.} {\bf#1} (#2) #3.}
\def\np#1#2#3   {{\em Nucl. Phys.} {\bf#1} (#2) #3}
\def\pcps#1#2#3 {{\em Proc. Cam. Phil. Soc.} {\bf#1} (#2) #3.}
\def\pl#1#2#3   {{\em Phys. Lett.} {\bf#1} (#2) #3}
\def\prep#1#2#3 {{\em Phys. Rep.} {\bf#1} (#2) #3}
\def\prev#1#2#3 {{\em Phys. Rev.} {\bf#1} (#2) #3}
\def\prl#1#2#3  {{\em Phys. Rev. Lett.} {\bf#1} (#2) #3}
\def\prs#1#2#3  {{\em Proc. Roy. Soc.} {\bf#1} (#2) #3.}
\def\ptp#1#2#3  {{\em Prog. Th. Phys.} {\bf#1} (#2) #3.}
\def\ps#1#2#3   {{\em Physica Scripta} {\bf#1} (#2) #3.}
\def\rmp#1#2#3  {{\em Rev. Mod. Phys.} {\bf#1} (#2) #3}
\def\rpp#1#2#3  {{\em Rep. Prog. Phys.} {\bf#1} (#2) #3.}
\def\sjnp#1#2#3 {{\em Sov. J. Nucl. Phys.} {\bf#1} (#2) #3}
\def\spj#1#2#3  {{\em Sov. Phys. JEPT} {\bf#1} (#2) #3}
\def\spu#1#2#3  {{\em Sov. Phys.-Usp.} {\bf#1} (#2) #3.}
\def\zp#1#2#3   {{\em Zeit. Phys.} {\bf#1} (#2) #3}

\title{\vspace{1cm}
\bf{ The proton and the photon, who is probing who? }
\vspace{2cm}}

\author{
 {\bf Aharon Levy} \\ 
{\small \sl School of Physics and Astronomy}\\ {\small \sl Raymond and 
Beverly Sackler Faculty of Exact Sciences}\\
  {\small \sl Tel--Aviv University, Tel--Aviv, Israel}
}
\date{ }
\maketitle

\vspace{5cm}

\begin{abstract}
  It is shown that by assuming Gribov factorization to hold at low $x$
  one obtains a simple relation between the structure function of the
  proton and that of the photon. By interpreting an observed structure
  in deep inelastic scattering (DIS) of leptons on protons as
  belonging to the proton, one can relate it to the structure of the
  exchanged photon, and visa versa. Predictions are given for the
  structure function of real and virtual photons at low $x$ by using
  data on the proton structure function together with the Gribov
  factorization relations.
\end{abstract}

\vspace{-22cm}
\begin{flushright}
TAUP 2406-97 \\
January 1997 \\
\end{flushright}

\setcounter{page}{0}
\thispagestyle{empty}
\newpage  

\section{Introduction}

Deep inelastic scattering (DIS) of leptons on protons are used as a
tool to gain information about the structure of the proton. A neutral
current (NC) reaction of the type $e p \to e X$ is interpreted naively
as a process in which the electron radiates a virtual gauge boson,
$\gamma$ or $Z^0$, which probes the structure of the proton. For low
enough virtualities, expressed by the negative square of the four
momentum transfer at the lepton vertex $Q^2$, one can neglect the
$Z^0$ exchange. Thus we say that the virtual photon $\gamma^*$ is
`looking' at the proton and probing its structure. There is no
question as to who `looks' at who and thus if some structure is
observed, it is attributed to the proton. The reason one does so is
because one assumes that $\gamma^*$ has no structure. How well is this
assumption justified?

Real photons are known to acquire structure when interacting with a
proton. A photon fluctuates into a $q \bar{q}$ pair and as long as the
fluctuation time $t_f$ is much larger than the interaction time
$t_{int}$ the photon interacts with the proton through the $q \bar{q}$
pair~\cite{ioffe}. A photon of energy $E_\gamma$ has a fluctuation
time of:
\begin{equation}
t_f \approx \frac{2E_\gamma}{m^2_{q\bar{q}}}
\end{equation}
while the interaction time is determined by the radius $r_p$ of the
proton: $t_{int} \sim r_p$.  A virtual photon of virtuality $Q^2$ has
a much shorter fluctuation time given by:
\begin{equation}
t_f \approx \frac{2E_\gamma}{m^2_{q\bar{q}} + Q^2}
\end{equation}
and thus does not have enough time to build up a structure before
interacting with the proton. 

However at small $x$ (the Bjorken scaling variable), the simple
picture of DIS is complicated by the long chain of gluon and quark
ladders which describes the process in Quantum Chromodynamics
(QCD)~\cite{roberts}. In this long chain of partons along the ladder,
where does one draw the line? Does one study the structure of the
proton? of the photon? of both? Does it make at all sense to speak
about the structure of virtual photons after the argument mentioned in
the earlier paragraph where it was shown that the fluctuation time
decreases with $Q^2$? That argument holds for the large $x$ region. At
low $x$ the fluctuation time becomes approximately $Q^2$ independent:
\begin{equation}
t_f \approx \frac{1}{2 m_p x}
\end{equation}
where $m_p$ is the proton mass and one assumes~\cite{afs}
$m^2_{q\bar{q}} \approx Q^2$. Thus for the low--$x$ region at HERA,
one expects also highly virtual photons to acquire some structure
before interacting with the proton. Thus we are back to the question
of how to interpret the DIS measurements. Are we measuring the proton
structure function $F_2^p$ or the photon structure function
$F_2^{\gamma^*}$? Who is probing who?
  
It is clear that physics can not be frame dependent~\cite{bj94}.  Thus
it must be that both descriptions are correct and reflect the fact
that cross sections are Lorenz invariant but time development is
not~\cite{lonya}. This means that it shouldn't matter whether one
interprets the cross section measurements as yielding the proton or
the photon structure function. By extracting one of them from the
cross section measurement, there should be a relation allowing to
obtain the other. On the other hand we know that at least as far as a
real photon is concerned, its structure function behaves very
differently from that of the proton one. As an example one can mention
the $Q^2$ scaling violation which is positive in the photon case for
all values of $x$, while for the proton they change from positive to
negative scaling violations as one moves to higher $x$
values~\cite{hawar}.
  
The purpose of this note is to suggest a way of relating the proton
and the photon structure function in the low--$x$ region, where the
interpretation of the results can be ambiguous. By extending the
Gribov factorization~\cite{gribov-fact} from the real photon to the
virtual photon case one can obtain $F_2^\gamma$ and $F_2^{\gamma^*}$
from the measured $F_2^p$ in the low--$x$ region.

\section{Gribov factorization}

Gribov factorization is based on the assumption that at high energies
the total cross section of two interacting particles is determined by
the property of the universal pomeron trajectory. This implies
relations between total cross section of various particles.  Gribov
factorization can be used~\cite{gribov-fact,rosner} to relate the
total $\gamma \gamma$ cross section, $\sigma_{\gamma \gamma}$, with
that of photoproduction, $\sigma_{\gamma p}$, and that of $p p$,
$\sigma_{p p}$, all at the same center of mass energy squared $W^2$:
\begin{equation}
\sigma_{\gamma \gamma}(W^2) = \frac{\sigma_{\gamma p}^2(W^2)}
{\sigma_{p p}(W^2)}.
\label{eq:fac}
\end{equation}
This relation is approximately borne out~\cite{sig-gg} with the
available measurements of $\sigma_{\gamma \gamma}$.

In case one of the photons is virtual, and assuming that Gribov
factorization is applicable also for virtual photons, one can write:
\begin{equation}
\sigma_{\gamma^* \gamma}(W^2,Q^2) = \frac{\sigma_{\gamma^* p}(W^2,Q^2)
\cdot\sigma_{\gamma p}(W^2)}{\sigma_{p p}(W^2)}.
\label{eq:fac*}
\end{equation}

The interest in the low--$x$ region stems from the fact that low $x$
for a given $Q^2$ value means high center of mass energy $W$, since
they are related through:
\begin{equation}
W^2 = Q^2 \left( \frac{1}{x} - 1 \right) - m_p^2 \simeq \frac{Q^2}{x}.
\label{eq:x-w2}
\end{equation}
In this region one can connect the proton structure function $F_2^p$
with the total $\gamma^* p$ cross section:
\begin{eqnarray} 
\sigma_{\gamma^* p}(W^2,Q^2) &=& \frac{4 \pi^2 \alpha}{Q^2} 
\frac{1}{1 - x} \left(1 + \frac{4 m_p^2 x^2}{Q^2} \right ) F_2^p(x,Q^2) \\
 &\approx & \frac{4 \pi^2 \alpha}{Q^2}F_2^p(x,Q^2). 
\end{eqnarray}
Similarly one can relate the photon structure function $F_2^\gamma$ to
the total $\gamma^* \gamma$ cross section:
\begin{equation}
\sigma_{\gamma^* \gamma}(W^2,Q^2) \approx 
\frac{4 \pi^2 \alpha}{Q^2}F_2^\gamma(x,Q^2).
\end{equation}
The approximate signs are well justified for the low--$x$ region. Using
equation~(\ref{eq:fac*}) one gets:
\begin{equation}
F_2^\gamma(x,Q^2) = F_2^p(x,Q^2) \frac{\sigma_{\gamma p}(W^2)}
{\sigma_{p p}(W^2)}.
\label{eq:fac-f2}
\end{equation}
This last equation connects the proton and the real photon structure
function at low $x$. By measuring one of them, the other can be
determined through relation~(\ref{eq:fac-f2}). Please note at this
point that $x$ should be treated as a measure of the center of mass
energy available in the interaction, as given in
equation~(\ref{eq:x-w2}).

The next step is to extend relation~(\ref{eq:fac*}) to the case where
both photons are virtual, one with virtuality of $Q^2$ and the other
with $P^2$. Assuming Gribov factorization to hold also for this case,
one can write:
\begin{equation}
\sigma_{\gamma^* \gamma^*}(W^2,Q^2,P^2) = \frac{\sigma_{\gamma^* p}(W^2,Q^2)
\cdot\sigma_{\gamma^* p}(W^2,P^2)}{\sigma_{p p}(W^2)}.
\label{eq:fac**}
\end{equation}
Just like above, we can replace some of the cross sections with structure 
functions to obtain:
\begin{equation}
F_2^{\gamma^*}(x,Q^2,P^2) = F_2^p(x,Q^2) \frac{\sigma_{\gamma^* p}(W^2,P^2)}
{\sigma_{p p}(W^2)}.
\label{eq:fac-f2*}
\end{equation}
This can also be expressed as:
\begin{equation}
F_2^{\gamma^*}(x,Q^2,P^2) = \left(\frac{4\pi^2\alpha}{P^2}\right)
\frac{F_2^p(x,Q^2)\cdot F_2^p(x,P^2)}{\sigma_{p p}(W^2)}.
\label{eq:fac-f2qp}
\end{equation}
Once again we obtained at low $x$ a relation between the measured proton 
structure function and that of the virtual photon. Thus 
equations~(\ref{eq:fac-f2}) and~(\ref{eq:fac-f2qp}) suggest that 
indeed at low $x$ both interpretation of the DIS cross sections 
as describing the structure of the proton or that of the photon are correct.

\section{Expected $W$ dependence of cross sections}

\subsection{The total $\gamma \gamma$ cross section, 
$\sigma_{\gamma \gamma}(W^2)$}

Donnachie and Landshoff (DL)~\cite{dl} have shown that all data on
hadron hadron cross sections can be expressed in the Regge picture as
a sum of two terms coming from contributions of the pomeron and the
reggeon trajectories. The contribution of the pomeron term is
proportional to $(W^2)^\Delta$ where the pomeron intercept is
expressed as $\alpha_P(0) = 1 + \Delta$. By fitting all available
total cross section data, they obtained the value $\Delta = $ 0.08.
The measurements at HERA of the total photoproduction cross
section~\cite{hera-sigtot} showed that this value of $\Delta$ can
explain also $\gamma p$ cross sections. Assuming that the pomeron term
dominates, $\sigma_{\gamma p}(W^2) = 0.068 (W^2)^{0.08}$ and
$\sigma_{p p}(W^2) = 21.7 (W^2)^{0.08}$, where both coefficients are
such that the resulting cross section is in milibarns and $W$ is in
GeV. By using relation~(\ref{eq:fac}) one obtains~\cite{ss}:
\begin{equation}
\sigma_{\gamma \gamma}(W^2) = 0.21\times10^{-3} (W^2)^{0.08},
\end{equation}
where again the cross section is expressed in milibarns. This
expression thus indicates that the energy behaviour of the total
$\gamma \gamma$ cross sections is the same as any other hadron hadron
or photoproduction cross section behaviour.

\subsection{The total $\gamma^* \gamma$ cross section,
$\sigma_{\gamma^* \gamma}(W^2,Q^2)$}

Equation~(\ref{eq:fac*}) can be used to get the relation:
\begin{equation}
\sigma_{\gamma^* \gamma}(W^2,Q^2) = 3.1\times10^{-3} 
\sigma_{\gamma^* p}(W^2,Q^2),
\end{equation}
which predicts that the $\gamma^* \gamma$ total cross sections behaves
at large energies in the same way as the $\gamma^* p$ cross section.
The HERA data~\cite{hera} can be well described by $\sigma_{\gamma^*
  p}(W^2,Q^2) \sim (W^2)^{\Delta(Q^2)}$~\cite{al-eps}, and thus one
expects also:
\begin{equation}
\sigma_{\gamma^* \gamma}(W^2,Q^2) \sim (W^2)^{\Delta(Q^2)}.
\end{equation}
Since $\Delta(Q^2)$ increases with $Q^2$ from the value of 0.08 at
$Q^2 \sim$ 0 to a value of about 0.3 at $Q^2 \sim$ 20 GeV$^2$, we
expect that also the $\gamma^* \gamma$ total cross section will have a
steeper $W$ behaviour as $Q^2$ increases.

\subsection{The total $\gamma^* \gamma^*$ cross section, 
$\sigma_{\gamma^* \gamma^*}(W^2,Q^2,P^2)$}

In the case where both photons are virtual the energy behaviour will
depend on both $Q^2$ and $P^2$. Using equation~(\ref{eq:fac**}), one
gets:
\begin{equation}
\sigma_{\gamma^* \gamma^*}(W^2,Q^2,P^2) \sim 
(W^2)^{\Delta(Q^2)+\Delta(P^2)-\Delta(0)}.
\end{equation}
Since $\Delta(P^2) > \Delta(0)$, this means that the $\gamma^*
\gamma^*$ cross section is expected to have a steeper $W$ dependence
than the total $\gamma^* p$ one.

\section{Comparison with existing data}

We compare the Gribov factorization relations to data from two
experiments which performed a double tag measurement and obtained $
\sigma_{\gamma^* \gamma^*}(W^2,Q^2,P^2)$. One is the TPC
collaboration~\cite{tpc} which tagged both photons with a virtuality
of $Q^2 = P^2 =$ 0.3 GeV$^2$. The other is the PLUTO
collaboration~\cite{pluto} which tagged one photon at $Q^2 =$ 5
GeV$^2$ and the other at $P^2 =$ 0.35 GeV$^2$.

\begin{figure}[h]
\begin{center}
  \includegraphics [bb=15 78 534 729,width=\hsize,totalheight=12cm]
  {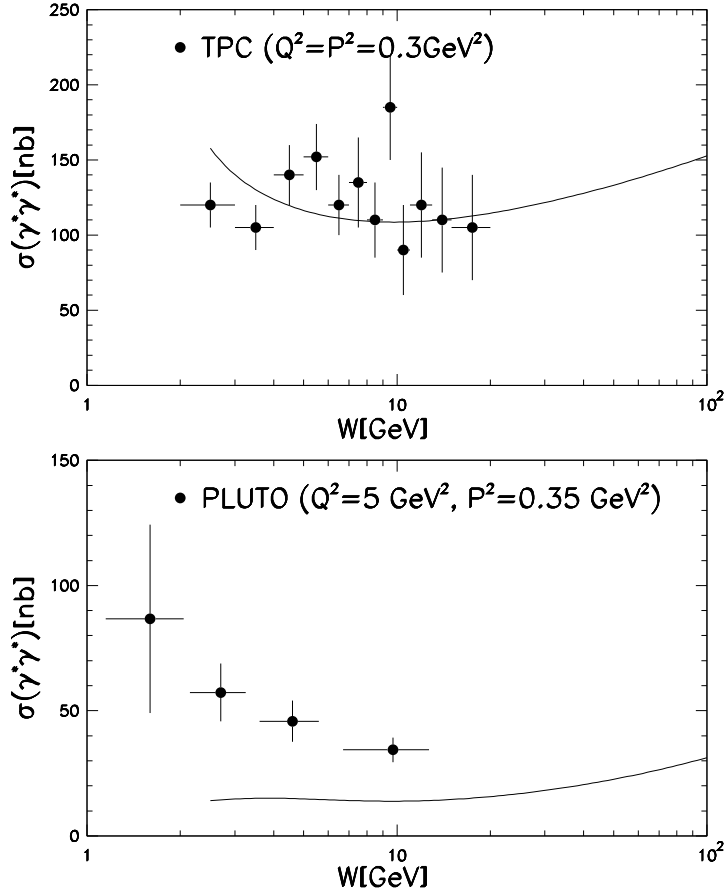}
\end{center}
\caption
{ The total $\gamma^* \gamma^*$ cross section as function of $W$ for
  the TPC data ($Q^2 = P^2 =$ 0.3 GeV$^2$) (upper plot) and for the
  PLUTO data ($Q^2 =$ 5 GeV$^2$, $P^2 =$ 0.35 GeV$^2$) (lower plot).
  The curves are the expectations from the Gribov factorization
  relation.
  }
\label{fig:sggvirt}
\end{figure}
Figure~\ref{fig:sggvirt} presents the total cross section data as
function of $W$. In order to calculate the expectations coming from
the Gribov factorization relation~(\ref{eq:fac**}) we use the DL
parameterization of the $p p$ total cross section. For the $\gamma^*
p$ cross section we use the ALLM~\cite{allm} parameterization which
gives a good description of all the measured data in the $Q^2$ range
between 0.3--2000 GeV$^2$. The curves appearing in the figure are
obtained using the above parameterizations.  Since the relations from
the Gribov factorization are expected to be valid in the low--$x$
region, the value of $W$ at which it can be applied is $Q^2$
dependent. If we call `low--$x$' as $x \sim 10^{-2}$, we expect the
factorization to work for $W \sim $ 5 GeV for $Q^2 \sim $ 0.3 GeV$^2$
and $W \sim$ 22 GeV for $Q^2 \sim$ 5 GeV$^2$. As can be seen from the
figure, the TPC data are in the range of validity of the Gribov
factorization and good agreement between the expectations and the data
is obtained. The PLUTO data are at lower $W$ values, where one can
still see a drop with energy toward the place where it would start
rising again.

\section{The photon structure function}

\subsection{Real photon structure function, $F_2^\gamma$}

\begin{figure}[h]
\begin{center}
\includegraphics
  [bb=15 88 534 732,width=\hsize,totalheight=12cm]
  {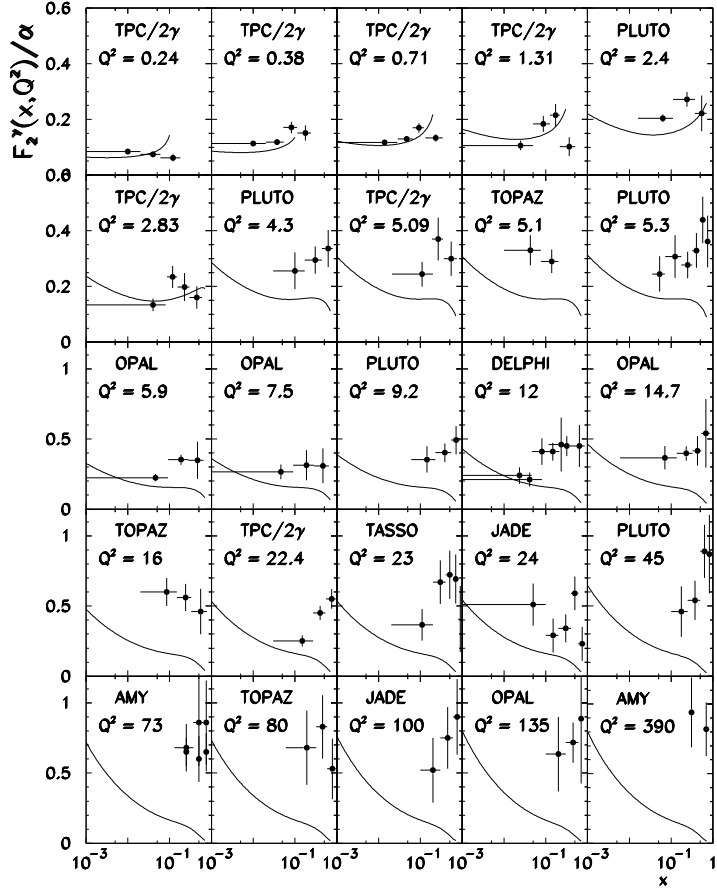}
\end{center}
\caption
{ A compilation of the data on the real photon structure function,
  $F_2^\gamma/\alpha$, as a function of $x$ for fixed $Q^2$ values.
  The curves are the expectations from the Gribov factorization
  relation.  }
\label{fig:f2gg-data-fac}
\end{figure}

The structure function of a real photon was measured in a $Q^2$ range
of 0.24--390 GeV$^2$. Most measurements, except those in the low $Q^2$
region, are in the relatively high--$x$ region. The data are plotted in
figure~\ref{fig:f2gg-data-fac} as function of $x$ for fixed $Q^2$
values.
In general, the structure function values seem to be decreasing as $x$
gets smaller. There is a clear lack of data in the region $x <$ 0.1.
The lines in the figure are the result of using the Gribov
factorization relation~(\ref{eq:fac-f2}). Instead of using the
available data on $F_2^p$, $\sigma_{\gamma p}$, $\sigma_{p p}$ and
extrapolate/interpolate the values to the kinematic region where the
$\gamma \gamma$ data is measured, we used the DL parameterization for
the $\gamma p$ and $ p p$ cross sections and the ALLM parameterization
for $F_2^p$. With the DL parameterization one obtains a simple relation:
\begin{equation}
F_2^\gamma/\alpha = 0.43 F_2^p .
\label{eq:f2g-f2p}
\end{equation} 
In the low $Q^2$ region, where low--$x$ data are available for the
photon structure function, the lines agree with the data. At higher
$Q^2$ regions, the predictions at high $x$ fall below the data as
expected since in the large--$x$ region factorization is not expected
to hold. In the low--$x$ region, the shape of $F_2^\gamma$ should
follow exactly that of $F_2^p$ according to
equation~(\ref{eq:f2g-f2p}). The parameterization of the photon parton
distributions given in~\cite{ss-par} has this feature in the low--$x$
region.  It would be most desirable to get data on the photon
structure function in the low--$x$ region to confirm this relation.

\subsection{Virtual photon structure function, $F_2^{\gamma^*}$}  

\begin{figure}[h]
\begin{center}
\includegraphics
  [bb=15 88 534 732,width=\hsize,totalheight=10cm]
  {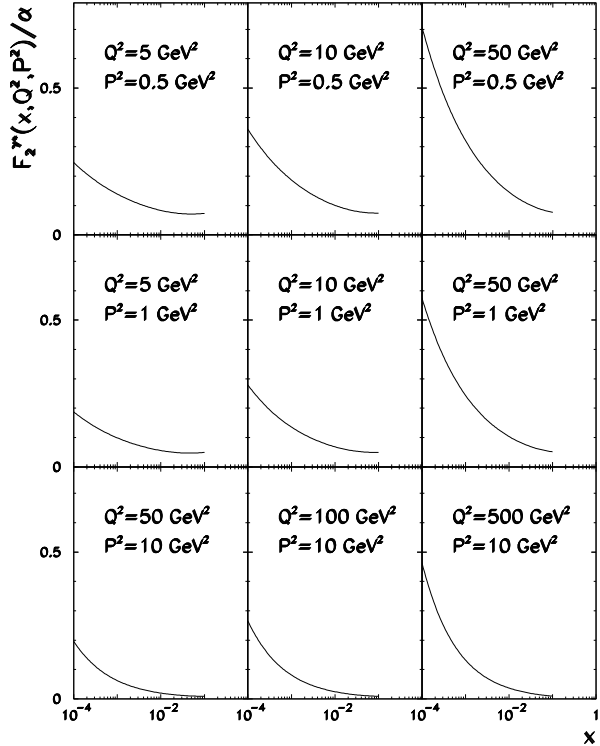}
\end{center}
\caption
{ The predicted behaviour of the virtual photon structure function as
  function of $x$, using the Gribov factorization relation, for values
  of $Q^2$ and $P^2$ as indicated in the figure.  
}
\label{fig:f2gvirt}
\end{figure}

There exist no measurement of $F_2^{\gamma^*}$ in the low--$x$ region.
We will use relation~(\ref{eq:fac-f2*}) to calculate the predictions
of the Gribov factorization for some specific $Q^2$ and $P^2$ values,
shown in figure~\ref{fig:f2gvirt}.
One sees a steep rise of the structure function with decreasing $x$
which is steeper than that observed for the proton structure function.
It is also interesting to note that even for a virtual photon of $P^2$
= 10 GeV$^2$, its structure function has a relatively high value at
low enough $x$ provided the `probing' photon is highly virtual
compared to the `probed' photon.

\section{Discussion and conclusions}

Gribov factorization has been applied for two cases where data exist and
in the region of its expected validity the results are consistent with
the data. One can use the measurements of $F_2^p$ at low $x$ in order
to predict $F_2^\gamma$ in that $x$ region. Predictions have been also
given for the structure function of virtual photons for some values of
$P^2$. 

The title of the paper indicated a problem of how to interpret at low
$x$ the results of the DIS experiments as to who is being probed by
who. Since in DIS one measures cross sections, in principle this
question should not matter since the result cannot be frame dependent.
This however means that by interpreting the result as coming from the
structure of the proton, one should be able to learn about the
structure of the photon. We showed here that if one can assume Gribov
factorization to hold also for the case of a virtual photon, one can
obtain a simple relation between the proton and the photon structure
function. Thus by measuring one of them, the other is determined
through the factorization relation.

The simple relation between the photon and proton structure function
obtained here relies on the extension of Gribov factorization to the
case of a virtual photon. This implies a factorizable pomeron also in
DIS processes, which so far seems to be borne out experimentally at
HERA~\cite{f2d3}. Experimental verification of the relations between
the photon and proton structure functions given above would be another
support for the factorizable pomeron in DIS. However it should be
noted that factorization breaking of the pomeron in DIS would mean
that the relations between the proton and photon structure function
are not as simple as those obtained in this paper.

We can conclude that at large $x$ the photon is probing the proton. At
low $x$ they seem to talk to each other: probing one teaches us about
the other.

\section*{Acknowledgments}
 
It is a pleasure to acknowledge fruitful discussions with 
H.~Abramowicz, J.~D.~Bjorken, L.~L.~Frankfurt, E.~G.~Gurvich and M.~Krawczyk.

\noindent This work was partially supported by the German--Israel 
Foundation (GIF).

%--------1---------2---------3---------4---------5---------6---------7---------

\end{document}